\documentclass[12pt]{iopart}

\usepackage{amssymb}

\begin{document}

\title{Finite-Size Corrections for Coulomb Systems in the
Debye--H\"uckel Regime}
\author{Aldemar Torres and Gabriel T\'ellez}

\address{Departamento de F\'{\i}sica\\
Universidad de Los Andes\\
A.A. 4976\\
Bogot\'a, Colombia\\
}

\eads{
\mailto{ald-torr@uniandes.edu.co},
\mailto{gtellez@uniandes.edu.co}
}

\begin{abstract}

It has been argued that for a finite two-dimensional classical Coulomb
system of characteristic size $R$, in its conducting phase, as
$R\rightarrow \infty $ the total free energy (times the inverse
temperature $\beta$) admits an expansion of the form: $\beta
F=AR^{2}+BR+\frac{1}{6}\chi \ln R,$ where $\chi $ is the Euler
characteristic of the manifold where the system lives. The first two
terms represent the bulk free energy and the surface free energy
respectively. The coefficients $A$ and $B$ are non-universal but the
coefficient of $\ln R$ is universal: it does not depend on the detail
of the microscopic constitution of the system (particle densities,
temperature, etc...). By doing the usual Legendre transform this
universal finite-size correction is also present in the grand
potential. The explicit form of the expansion has been checked for
some exactly solvable models for a special value of the coulombic
coupling. In this paper we present a method to obtain these
finite-size corrections at the Debye--H\"uckel regime. It is based on
the sine-Gordon field theory to find an expression for the grand
canonical partition function in terms of the spectrum of the Laplace
operator. As examples we find explicitly the grand potential expansion
for a Coulomb system confined in a disk and in an annulus with ideal
conductor walls.

\end{abstract}

\pacs{05.20.Jj, 51.30.+i}


\maketitle

\section{Introduction}

There are several reasons to study models over a finite-size
region. For instance, with the recent advance of computers, much
information on statistical models has been derived from computer
simulations, which are necessarily limited to systems of
finite-size. Also experimental systems are finite (although very
large). The finite scaling hypothesis allows the study of some
response functions for such finite-size models. For a $d$-dimensional
system, this finite scaling hypothesis states that if a given response
function (for example the susceptibility in a magnetic system)
diverges in the bulk like $\xi ^{d-2x}$, where $\xi$ is the
correlation length and $x$ is called the scaling dimension of the
corresponding studied quantity (the magnetization in the above
example), then in a finite system of characteristic size $R$, the
response function should obey the scaling law $R^{d-2x}\Phi (R/\xi )$,
where $\Phi $ is some universal scaling function~\cite{Bergersen,
cardy chap}. At the critical point where the correlation length
diverges, the response function is then proportional to
$R^{d-2x}$. However the scaling of the free energy at criticality is
less well understood at least for arbitrary dimension. In two
dimensions, using methods from conformal field theory, it has been
shown that for a finite system with smooth boundary, of characteristic
size $R$ as $R$ $\rightarrow \infty$, at criticality, the total free 
energy $F$ has a large-$R$ expansion of the form~\cite{cardy chap, cady
finite size}
\begin{equation}
\label{eq:F-finite-size}
\beta F=AR^{2}+BR-\frac{c\chi }{6}\ln R+\mathcal{O}(1)
\ .
\end{equation}
with $\beta=(k_B T)^{-1}$ the inverse reduced temperature. The first
two terms represent respectively the bulk free energy and the
``surface'' (perimeter in two dimensions) contribution to the free
energy. In general, the coefficients $A$ and $B$ are non universal but
the dimensionless coefficient of $\ln R$ is universal depending only
on $c$, the conformal anomaly number, and on $\chi$, the Euler
characteristic of the manifold ($\chi =2-2h-b$, where $h$ is the
number of handles and $b$ is the number of boundaries). Surprisingly
enough, for classical Coulomb systems in their conducting phase ---not
at criticality--- this expansion for the free energy seems to holds
with $c=1$ and a change of sign in the last term, and it has been
explicitly checked for Coulomb systems lying on some simple geometries
for some exactly solvable models for the fixed temperature defined by
$\beta q^{2}=2$ where $\beta ^{-1}=k_{B}T$ and $\pm q$ are the charges
of the particles~\cite{Finite zise jancov,Finite zise
Gabriel,dosdesfera,Tellez-tcp-disque-neumann} and also it has been
verified numerically for the one-component plasma confined in a
disk~\cite{Tellez-Forrester-2dOCP-Gamma=4-6} for other values of the
coulombic coupling. The existence of the universal finite-size
correction has also been proved for the two-component plasma confined
in a disk with hard walls in the whole regime where the system of
point particles is stable ($\beta
q^2<2$)~\cite{Samaj-Janco-density-corr-TCP}. For the one-component
plasma~\cite{Janco-Trizac-sphere-correction-ocp}, for the symmetric
two-component plasma~\cite{Janco-sphere-correction-tcp} and for the
asymmetric two-component
plasma~\cite{Samaj-asym-tcp-correction-sphere} confined in a sphere
the existence of the finite-size correction has been proved for any
temperature (provided that the system is stable) by application of the
stereographic projection and some non-trivial sum rules concerning the
density-density correlation function in the plane
geometry~\cite{sum-rule-density-density,sum-rule-Kalinay-sixth-moment}.

Although the particle
and charge correlations in Coulomb systems in their conducting phase
are short-ranged because of the screening, the electric potential
correlations are long
ranged~\cite{Lebo-Martin,Janco-screen-correl-revisit}.  It has been
argued~\cite{Finite zise jancov,Finite zise Gabriel} that in this
sense the system is comparable to a critical system and therefore the
expansion of the free energy~(\ref{eq:F-finite-size}) should hold.

In this paper we present a method to obtain the grand potential
finite-size expansion for a confined Coulomb system in the
Debye--H\"uckel regime and we check the existence of the logarithmic
universal finite-size correction. The Debye--H\"uckel regime is
defined by the requirement that the average coulombic energy is much
smaller than the thermal energy~\cite{Debye Screening,Debye screening
rigurous}. By the usual Legendre transform between the free energy and
the grand potential it can be inferred that the free energy will have
the same logarithmic finite-size corrections as the grand
potential. Our method is based on the sine--Gordon field
theory~\cite{Sine gordon} to calculate the grand canonical partition
function.

\section{Sine-Gordon theory in the Debye--H\"uckel regime}
\label{sec:method}

There is a well-known analogy between statistical mechanics and
quantum field theory: often the partition function of a
$d$-dimensional statistical model is formally analogous to the
generating functional of a quantum field in $d$ space-time dimensions
in the Euclidean formalism~\cite{conformal}. The simplest example of a
quantum field theory which has a relevance in statistical mechanics is
the free Boson or Gaussian model. In this section we show that the
grand canonical partition function of a Coulomb system in the
Debye--H\"uckel regime may be represented as the generating functional
of a massive free Boson theory and therefore it can be obtained
exactly from a Gaussian integration as an infinite product of
functions of the eigenvalues of the Laplace operator calculated on the
manifold where the system lives.

Let a classical (i.~e.~non-quantum) Coulomb system be composed of
$\alpha =1,\ldots,r$ species of particles each of which have
$N_{\alpha }$ charges $q_{\alpha }$ confined in a $d$-dimensional
Riemannian manifold of volume $V$. Suppose that the system is confined
by grounded ideal conductor walls, thus imposing vanishing Dirichlet
boundary conditions to the electric potential. We shall describe the
system in the grand canonical ensemble with fugacities
$\zeta_{\alpha}=e^{\beta\mu_{\alpha}}/\Lambda^d$ for the species
$\alpha$, where $\mu_{\alpha}$ is the chemical potential and $\Lambda$
is the de Broglie thermal wavelength which appears when the Gaussian
integration over the kinetic part of the hamiltonian is carried
out. For a finite but macroscopically large system, the interior of
the system will be at an almost constant electric potential
$\psi_0$. The value of $\psi_0$ is controlled by the choice of the
fugacities. We will suppose in the following that the fugacities
satisfy the relation
\begin{equation}
\label{eq:pseudo-neutrality0}
\sum_{\alpha}
q_{\alpha} \zeta_{\alpha}=0
\end{equation}
which is often referred in the literature~\cite{review} as the
pseudo-neutrality condition. In the~\ref{sec:appendix-B} we consider
the general case when the fugacities do not satisfy the
condition~(\ref{eq:pseudo-neutrality0}).

Let us introduce the Coulomb potential for a non-confined system for
unit charges
\begin{equation}
v_d^0(\mathbf{r},\mathbf{r}')
=
\cases{
\frac{1}{\left|
\mathbf{r}- \mathbf{r}'\right|}\,,
&if $d=3$\\
-\ln  \frac{\left|\mathbf{r}
-\mathbf{r}'\right|}{L}  \,,
& if $d=2$.
}
\end{equation}
In two dimensions, a solution of Poisson equation that vanishes at
large distances does not exist, therefore it is necessary to introduce
an arbitrary length $L$ that fixes the zero of the electric
potential. However as we will see later, in the formulation of
Debye--H\"uckel theory proposed here it will be necessary to suppose
that $L\to\infty$, thus receding the zero of the electric potential to
infinity. The necessity to choose $L\to\infty$ also appears in the
formulation of Debye--H\"uckel theory from Ornstein--Zernicke equation
and the approximation of the direct correlation function by the
Coulomb potential~\cite{jancov l inf}.

The Coulomb potential in $d$ dimensions for the system confined with
Dirichlet boundary conditions will be referred as $v_d$.  This
potential satisfy the Poisson equation
\begin{equation}
\Delta v_{d}(\mathbf{r},\mathbf{r}')=-s_{d}\delta
(\mathbf{r}-\mathbf{r}')  
\end{equation}
with \ $s_{2}=2\pi $ and\ $s_{3}=4\pi $, and the Dirichlet boundary
condition. If one considers $v_d(\mathbf{r},\mathbf{r}')$ as the
kernel of an integral operator which we will also call $v_d$, we have
$v_{d}=-s_{d}{\Delta }^{-1}$.  Let $\Psi_{n}(\mathbf{r})$ be the
normalized eigenfunctions of the Laplacian with Dirichlet boundary
conditions, that is $\Delta
\Psi_{n}(\mathbf{r})=\lambda_{n}\Psi_{n}(\mathbf{r})$ where
$\lambda_{n}\leq 0$ are the corresponding eigenvalues. These functions
are also eigenfunctions of $v_d$ with the corresponding eigenvalues
$-s_d/\lambda_n\geq 0$. Consider two particles located at
$\mathbf{r}_{\alpha,i}$ and $\mathbf{r}_{\beta,j}$. A standard
operator spectral decomposition gives for the interparticle potential
\begin{eqnarray}
v_{d}(\mathbf{r}_{\alpha,i},\mathbf{r}_{\beta,j})
&=&-
\sum_{n} \frac{s_{d}}{\lambda_{n}}
\,
\overline{\Psi _{n}(\mathbf{r}_{\alpha,i})}
\Psi _{n}(\mathbf{r}_{\beta,j})\,.
\label{iterp}
\end{eqnarray}
The bar over $\Psi$ indicates complex conjugation. Additionally to the
interparticle energy we consider the energy of each particle located
at $\mathbf{r}_{\alpha,i}$ in presence of the field produced by itself
$v^{0}_d(\mathbf{r}_{\alpha,i},\mathbf{r}_{\alpha,i})=
v_{S-E}(\mathbf{r}_{\alpha,i})$. This term is (twice) the self-energy
of a unit charge. Proceeding as in (\ref{iterp}) it may be given by
\begin{equation}
v_{S-E}(\mathbf{r}_{\alpha,i})=
-
\sum_{k}\frac{s_{d}}{\lambda _{k}^{0}}\left|
\Psi _{k}^{0}(\mathbf{r}_{\alpha,i})\right| ^{2}
\label{selfenergy operator}
\end{equation}
here the $\lambda _{k}^{0}$ refer to eigenvalues calculated for the
system without boundaries. Of course, because of the form of the
Coulomb potential the self-energy is in fact infinite. This divergence
may be avoided by the introduction of a short distance
potential~\cite{Debye screening rigurous} to cutoff the singularity of
the Coulomb potential at the origin. To simplify the notation, we will
not write down explicitly this short distance potential in what
follows. It should be noted however that the introduction of this
short distance potential is mandatory in classical statistical
mechanics of Coulomb systems in order to have a well defined partition
function in three dimensions (in two dimensions at low coulombic
couplings a system of point particles is stable).  On the other hand
it turns out that the Debye--H\"uckel approximation is well defined
for a system of point particles: as we shall see later the short
distance potential does not appear in the final results.  Let us
remark that for systems governed by quantum mechanics if all particles
of a same sign are fermions then the system is
stable~\cite{Lieb-Lebowitz}. This is the case in nature where quantum
mechanics are responsible for the creation of stable bound
states. Therefore our classical analysis will apply only to fully
ionized systems.

The potential energy of the system is given by
\begin{equation}
\fl
H=\frac{1}{2}\sum_{\alpha ,\beta }\sum\nolimits_{i,j}^{^{\prime
}}q_{\alpha }q_{\beta }v_{d}(\mathbf{r}_{\alpha,i},
\mathbf{r}_{\beta,j})+\frac{1}{2}\sum_{\alpha}\sum_{i}
q_{\alpha}^2 \left[
v_d(\mathbf{r}_{\alpha,i},\mathbf{r}_{\alpha,i})
-v_{S-E}(\mathbf{r}_{\alpha,i})\right]
\,.
\end{equation}
The prime in the first summation mean that the case when$\ \alpha
=\beta $ and $i=j$ must be omitted. The first term is the usual
interparticle energy between pairs. The second term is the Coulomb
energy of a particle and the polarization surface charge density that
the particle has induced in the boundaries of the system. When the
method of images is applicable to compute the Coulomb potential $v_d$,
this energy can also be seen as the energy between each particle and
its image.

Using the microscopic charge density defined as
\begin{equation}
\label{eq:rho}
\hat{\rho}(\mathbf{r})=\sum_{\alpha =1}^{r}\sum_{i=1}^{N_{\alpha
}}q_{\alpha }\delta (\mathbf{r}-\mathbf{r}_{\alpha,i})  
\end{equation}
we can write the potential part of the Hamiltonian of the system as
\begin{equation}
H=\frac{1}{2}
\int_{V}d\mathbf{r}\int_{V}d\mathbf{r}'\,\hat{\rho}(\mathbf{r})
v_d(\mathbf{r},\mathbf{r}^{\prime })\hat{\rho}(\mathbf{r}^{\prime})
-\frac{1}{2}\sum_{\alpha =1}^{r}\sum_{i=1}^{N_{\alpha }}q_{\alpha
}^{2}v_{S-E}( \mathbf{r}_{\alpha,i}) \,.
\label{H}
\end{equation}
Notice that with this notation, in the first term, the terms
$q_{\alpha}^2 v_d(\mathbf{r}_{\alpha,i},\mathbf{r}_{\alpha,i})/2$ have
been included. Often in the sine-Gordon transformation~\cite{Sine
gordon,Kholodenko} the second term is omitted in the Hamiltonian in
equation~(\ref{H}), which implies that the self-energy (infinite for
point particles) is included in the Hamiltonian when it should not
be. This problem can be cured with a proper renormalization of the
fugacity~\cite{Samaj-TCP}, however this method is not the more
convenient to use for the Debye--H\"uckel approximation, therefore we
will proceed to subtract the self-energy from the start as shown in
equation~(\ref{H}).

Now, using the well-known Gaussian integral
\begin{equation}
e^{\frac{1}{2}\mathbf{B\cdot A}^{-1}\mathbf{\cdot B}}=
\frac{\int d\mathbf{X}\,
e^{-\frac{1}{2}\mathbf{X\cdot A\cdot X+B\cdot X}}}{\int d\mathbf{X}\,
e^{-\frac{ 1}{2}\mathbf{X\cdot A\cdot X}}} \label{gaux}
\end{equation}
we can represent the Boltzmann factor as\footnote{Rigorously speaking
  this Gaussian transformation can not be formulated with the Coulomb
  potential $v_d(\mathbf{r},\mathbf{r}')$ because this potential
  diverges when $\mathbf{r}=\mathbf{r}'$. This problem can be solved
  as in Ref.~\cite{Kennedy} using instead a potential like
  $(1-e^{-\kappa r/\varepsilon})/r$ which is regularized at short
  distances and taking the limit $\varepsilon\to0$ at the end of the
  calculations. Again, for simplicity, we will omit explicitly this
  detail in what follows. }
\begin{equation}
e^{-\beta H}=\left\langle \exp \left[ - \beta \int {i\hat{\rho}(\mathbf{r
})}\phi (\mathbf{r})\,d\mathbf{r}+\frac{\beta}{2} \sum_{\alpha
=1}^{r}\sum_{i=1}^{N_{\alpha }}q_{\alpha }^{2}v_{S-E}(\mathbf{r}_{\alpha
_{i}})\right] \right\rangle  \label{Boltzmann}
\end{equation}
where we have defined the average of any operator $\hat{o}$ as
$\left\langle \hat{o}\right\rangle =\frac{1}{Z}\int \mathcal{D}\phi\,
\hat{o}\,e^{\frac{\beta }{2s_{d}}\int \phi (\mathbf{r})\Delta
\phi (\mathbf{r})\,d\mathbf{r} }$, with $Z=\int \mathcal{D}\phi\,
e^{\frac{\beta }{2s_{d}}\int \phi (\mathbf{r}) \Delta \phi
(\mathbf{r})\,d\mathbf{r}}$. On the other side, using~(\ref{eq:rho})
and~(\ref{Boltzmann}) after some calculations the grand partition
function is given by~\cite{Sine gordon} 
\begin{eqnarray}
\Xi(\beta ,\zeta _{1},\ldots,\zeta _{r},V)
&=&\sum_{N_{1}=0}^{\infty } \cdots \sum_{N_{r}=0}^{\infty }\frac{
\zeta _{1}^{N_{1}}\cdots\zeta _{r}^{N_{r}}}{N_{1}!\ldots N_{r}!}\int \cdot
\cdot \cdot \int e^{-\beta H}\prod_{\alpha =1}^{r}\prod_{i=1}^{N_{\alpha }}d
\mathbf{r}_{\alpha,i}  \nonumber \\
&=&\left\langle \exp \left[ \sum_{\alpha }^{r}\zeta _{\alpha }\int
e^{\beta \left( -iq_{\alpha }\phi (\mathbf{r})+
\frac{q_{\alpha }^{2}}{2}v_{S-E}(\mathbf{r})\right) }
d\mathbf{r}\right] \right\rangle  \label{G}
\end{eqnarray}
with $V$ the volume of the manifold containing the system. We see from
equation (\ref{G}) that the partition function for a gas of particles
with two-body interactions may be obtained as the average of the
partition function of an ideal gas in an external fluctuating
potential $i\phi(\mathbf{r})$. In general the integration involving
the calculation of (\ref{G}) cannot be performed analytically ---with
the notable exceptions of the two-dimensional two-component plasma
(symmetric $1:1$ and asymmetric $+2:-1$) which has been exactly solved
in the bulk and in some semi-infinite
geometries~\cite{Samaj-TCP,Samaj-Janco-TCP-metal,Samaj-TCP-ideal-dielec,Samaj-TCP-asym}.

The coulombic coupling $\Gamma$ is defined as the average Coulomb
energy divided by the thermal energy. We can actually define a
coupling for each species of particles as follows. In two dimensions
these are defined as $\Gamma_{2,\alpha}=\beta q^{2}_{\alpha}$. On the
other hand, in three dimensions $\Gamma_{3,\alpha}=\beta q_{\alpha}^2
\zeta_{\alpha}^{1/3}$. In the Debye--H\"uckel regime we have
$\Gamma_{d,\alpha} \ll 1$ and we can expand
\begin{equation}
\fl
\label{eq:DH-approximation}
\exp\left[\beta \left[- i q_{\alpha }\phi
(\mathbf{r} )+\frac{q_{\alpha}^{2}}{2}v_{S-E}(\mathbf{r})\right]
\right]
=
1-i \beta q_{\alpha }\phi (\mathbf{r})+\frac{\beta
q_{\alpha}^{2}}{2} v_{S-E}(\mathbf{r})-\frac{\left( \beta q_{\alpha
}\phi \mathbf{(\mathbf{r})}\right) ^{2}}{2}
+o\left(\Gamma_{d,\alpha}^2\right)
.  
\end{equation}
In two dimensions the field $\phi(\mathbf{r})$ has dimensions of
charge $q_{\alpha}$ therefore the above approximation is an expansion
to the order $(\Gamma_{2,\alpha})^2$ in the coulombic couplings
$\Gamma_{2,\alpha}$. In three dimensions the field $\phi(\mathbf{r})$
has dimensions of charge / distance. One can do a change of variable
in the functional integral to have a dimensionless field
$\tilde{\phi}= \zeta^{-1/3} \phi/q$, then it is clear that the
approximation~(\ref{eq:DH-approximation}) is again an expansion to the
second order in the coulombic couplings $\Gamma_{3,\alpha}$. Notice
that the self-energy term $\beta q_{\alpha}^{2} v_{S-E}(\mathbf{r})/2$
is already of order $(\Gamma_{d,\alpha})^2$.  This can be seen by
noticing that the covariance of the Gaussian measure $\frac{1}{Z}
\mathcal{D}\phi\,e^{\frac{\beta }{2s_{d}}\int \phi \Delta \phi}$ is
$\langle \phi(\mathbf{r})\phi(\mathbf{r}')\rangle=\beta^{-1}
v_{d}(\mathbf{r},\mathbf{r}')$. Then it is clear that the self-energy
term $\beta q_{\alpha}^{2} v_{S-E}(\mathbf{r})/2$ is of the same order
as $(\beta q_{\alpha} \phi(\mathbf{r}))^2$.  Therefore we do not
include any terms of order $(\beta q_{\alpha}^{2}
v_{S-E}(\mathbf{r})/2)^2$ or $\beta^2 q_{\alpha}^3\phi(\mathbf{r})
v_{S-E}(\mathbf{r})$ which are of higher order in the
expansion~(\ref{eq:DH-approximation}).

Using equation~(\ref{eq:DH-approximation}) and the pseudo-neutrality
condition~(\ref{eq:pseudo-neutrality0}) we have
\begin{equation}
\fl
\Xi(\beta ,\zeta _{1}...\zeta _{\alpha },V)=\left\langle \exp 
\left[ - \int \sum_{\alpha }\frac{\zeta _{\alpha }\left( \beta
q_{\alpha }\phi \mathbf{(\mathbf{r})}\right) ^{2}}{2}\, d\mathbf{r}
\right] \right\rangle e^{-\sum_{\alpha =1}^{r}\sum_{n}\frac{\mathbf{\beta }
\zeta _{\alpha }q_{\alpha }^{2}s_{d}}{2\lambda _{n}^{0}}}e^{\sum_{\alpha
}V\zeta _{\alpha }}  \label{xi}
\end{equation}
where the spectral decomposition~(\ref{selfenergy operator}) of the
self-energy and the normalization condition $\int \left| \Psi
_{n}^{0}(\mathbf{r}_{\alpha })\right| ^{2}d\mathbf{r}=1$ have been
used to write the contribution the self-energy terms as a sum over the
eigenvalues $\lambda_n^0$ of the Laplacian without boundaries.  Now
performing the Gaussian integration the averaged quantity equals to
\begin{equation}
\label{eq:gaussian-inte}
\fl
\frac{1}{Z}\int \mathcal{D}\phi\, e^{ \frac{1}{2}\int \phi \mathbf{(
\mathbf{r})}\left( \frac{\beta {\Delta }}{s_{d}}-\sum_{\alpha
}\zeta _{\alpha }\left( \beta q_{\alpha }\right) ^{2}\right) \phi 
\mathbf{(\mathbf{r})}d\mathbf{\mathbf{r}}} =\left( \det \left[ 1-\frac{
\sum_{\alpha }s_{d}\zeta _{\alpha }\beta q_{\alpha }^{2}}{{\Delta 
}}\right] \right) ^{-1/2}
.
\end{equation}
Notice that the averaged quantity we had just computed is the
generating functional of a free Boson theory with mass proportional to
the inverse Debye length defined by $\kappa=\sqrt{\sum_{\alpha
}s_{d}\zeta _{\alpha }\beta q_{\alpha }^{2}}$. Using the invariance of
the determinant we obtain for the grand canonical partition function
\begin{equation}
\Xi(\beta ,\zeta _{1},\ldots,\zeta _{\alpha },V)=\left(
\prod_{m}\left( 1-\frac{\kappa^{2}}{\lambda _{m}}\right)
\prod_{n} e^{\frac{\kappa^{2}}{\lambda _{n}^{0}}} \right)
^{-1/2}e^{\sum_{\alpha }V\zeta _{\alpha }} \label{gp}
.
\end{equation}
This is simply a product of factors that are a function of the
eigenvalues $ \lambda _{i}$ of ${\Delta }.$ The $\lambda _{i}s$ depend
on the shape of the domain in which the Coulomb system lies. As we
see, they constitute a natural way to introduce the information on the
domain to calculate the corresponding finite-size expansion of the
grand potential $\Omega$. It is interesting to point out here that in
the case of a non-confined system $\lambda _{n}=\lambda _{n}^{0} $ and
the infinite products in (\ref{gp}) become a regularized Weierstrass
product
$\prod_{n}\left[\left(1-\frac{\kappa^{2}}{\lambda_{n}^0}\right)
e^{\kappa^{2}/\lambda _{n}^{0}}\right]$. The $\prod
e^{\kappa^{2}/\lambda _{n}^{0}}$ term cancels out the ultraviolet
divergence coming from of the $\prod
\left(1-\frac{\kappa^{2}}{\lambda_{n}}\right)$ term. This product
converges for systems in three dimensions~\cite{spectral
functions}. However, as it will be seen in the next section, in two
dimensions some infrared divergences appear and the product must be
regularized by introducing a lower cutoff. The sine-Gordon
transformation has been known for some time~\cite{Debye screening
rigurous, Sine gordon, Kholodenko}. For three dimensional non-confined
systems the sine-Gordon transformation have been used to go beyond the
Debye--H\"uckel approximation and to perform low
fugacity~\cite{Caillol-low-fugas}, high
temperature~\cite{Caillol-high-temp} or loopwise~\cite{Caillol-loop}
expansions. The main point of this section was to show that the proper
subtraction of the self-energy terms (which have to be added initially
to perform the sine-Gordon transformation) leads to a well-defined
expression for the grand potential in the Debye--H\"uckel
approximation which could be eventually evaluated for confined
systems. In the~\ref{sec:appendix} we show how this formulation of
Debye--H\"uckel theory is related to the usual one.  In the following
section we apply this method to the calculation of the grand potential
$\Omega =-k_{B}T\ln \Xi $ of a Coulomb system confined in some simple
geometries.

\section{Finite-size corrections to the grand potential for a
confined Coulomb system}
\label{sec:confined}

\subsection{Non-confined systems: the bulk}

Before applying our method to confined systems let us illustrate some
points of the calculation of the grand potential from
equation~(\ref{gp}) for a bulk system.  For a $d$-dimensional
non-confined system the wave functions corresponding to the $\lambda
_{k}^{0}$ are given by $\Psi _{\mathbf{k}}^{0}
(\mathbf{r})=\frac{e^{i\mathbf{k}\cdot \mathbf{r}}}{\sqrt{V}}$ \ and
$\Delta \Psi _{\mathbf{k}}^{0}(\mathbf{r})=\lambda _{k}^{0}\Psi
_{\mathbf{k}}^{0}(\mathbf{r})=-\mathbf{k}^{2}\Psi _{\mathbf{k}}^{0}(
\mathbf{r})$. Then replacing into equation~(\ref{gp}) the grand
potential is given by
\begin{equation}
  \label{eq:grand-potential-bulk}
\beta \Omega =\frac{V}{2}
\int \frac{d^{d}\mathbf{k}}{(2\pi )^{d}}
\left[ \ln \left( 1+\frac{\kappa^{2}}{k^{2}}\right) -
\frac{\kappa^{2}}{k^{2}}\right] -\sum_{\alpha }V\zeta _{\alpha }
\,.
\end{equation}
In three dimensions $d=3$ the above integral is convergent giving the
result for the bulk grand potential per unit volume~\cite{Kennedy}
\begin{equation}
  \label{eq:Omega-bulk-3D-fugacite}
  \frac{\beta\Omega}{V}=-\frac{\kappa^3}{12\pi}-\sum_{\alpha}
  \zeta_{\alpha}
\,.
\end{equation}
The density $n_{\alpha}$ of the particles can be obtained by using the
usual thermodynamic relation
\begin{equation}
  n_{\alpha}=-\zeta_{\alpha}
  \frac{\partial (\beta\Omega/V)}{\partial \zeta_{\alpha}}
  = \zeta_{\alpha} +\frac{\kappa}{2}  \beta
  q_{\alpha}^2\zeta_{\alpha}
\end{equation}
which can be replaced back into
equation~(\ref{eq:Omega-bulk-3D-fugacite}) to give the well-known
equation of state from Debye--H\"uckel theory~\cite{Kennedy,McQuarrie}
\begin{equation}
  \beta p=-\frac{\beta\Omega}{V}=\sum_{\alpha} n_{\alpha} -
  \frac{\kappa_{\mathrm{DH}}^{3}}{24\pi}
\,.
\end{equation}
Notice that in the last equation the Debye length that we have defined
by $\kappa^{-1}=\left[\sum_{\alpha} s_d \zeta_{\alpha}\beta
q_{\alpha}^2\right]^{-1/2}$ has been replaced by the usual Debye length in
terms of the density $\kappa_{\mathrm{DH}}^{-1}=\left[\sum_{\alpha} s_d
n_{\alpha}\beta q_{\alpha}^2\right]^{-1/2}$. This is correct at the order of
approximation we are working on since
$\kappa=\kappa_{\mathrm{DH}}[1+\mathcal{O}(\Gamma_{3,\alpha}^{3/2})]$.

Let us point out that the proper substraction of the self-energy terms
makes the integral~(\ref{eq:grand-potential-bulk}) convergent and
avoids the need to use some other arbitrary regularization scheme,
like for instance dimensional regularization used in
Ref.~\cite{Kholodenko} which by the way yield the incorrect result
$\sum_{\alpha}\zeta_\alpha -\kappa^3/(24\pi)$ for the pressure when it
is expressed in terms of the \textit{fugacities}. Our regularization
scheme is actually equivalent to the normal ordering for the product
$:\phi(\mathbf{r})^2:$ used in Ref.~\cite{Kennedy}.

As it was pointed out in the preceding section the infinite
product~(\ref{gp}) for a non-confined system is a regularized
Weierstrass product. The order of the sequence of the Laplacian
eigenvalues is $\mu=d/2$~\cite{spectral functions}, therefore for
$d=3$, $\mu=3/2>1$ with integer part equal to one, and the terms
$\exp(\kappa^2/\lambda_k^{0})$ in the product~(\ref{gp}) are enough to
regularize the infinite product.

The situation in two dimensions $d=2$ is more delicate since $\mu=1$
is a limiting case. If we blindly try to
compute~(\ref{eq:grand-potential-bulk}) we will notice that the
integral is not well-defined for $k\to 0$. Trying to cure an
ultraviolet divergence, we introduced an infrared one. The problem can
be traced back to the spectral decomposition~(\ref{iterp}) of the
Coulomb potential. Evaluating explicitly the interparticle energy
using the expression (\ref{iterp}) gives
\begin{equation}
v_{2}(\mathbf{r},\mathbf{r}^{\prime })=\frac{1}{2\pi }\int_{0}^{2\pi
}\int_{0}^{\infty }\frac{ke^{ik\left| \mathbf{r}-\mathbf{r}^{\prime
}\right| \cos \theta }}{k^{2}}dk d\theta =\int_{0}^{\infty
}\frac{dk}{k}J_{0}(k\left| \mathbf{r}-\mathbf{r}^{\prime }\right| )
\end{equation}
with $J_0$ the Bessel function of order 0. This integral diverges at
$k=0$. To avoid this, we introduce a cutoff $k_{\min}$ at
$k\rightarrow 0$
\begin{eqnarray}
v_{2}(\mathbf{r},\mathbf{r}^{\prime }) &=&
\int_{k_{\min }}^{\infty }\frac{dk}{k}J_{0}
(k\left| \mathbf{r}-\mathbf{r}^{\prime }\right| ) \\
&=& -C +\ln 
\frac{2}{\left| \mathbf{r}-\mathbf{r}^{\prime }\right| k_{\min}}
+o(1)
\label{eq:kmin-introduccion}
\end{eqnarray}
where $k_{\min }\rightarrow 0$ and $C$ is the Euler constant.  Since
we know that $v_{2}(\mathbf{r},\mathbf{r}^{\prime })=-\ln \frac{\left|
\mathbf{r}-\mathbf{r}^{\prime }\right| }{L}$ we can find the
expression for $ k_{\min }$ by comparison: $v_{2}(\left|
\mathbf{r}-\mathbf{r}^{\prime }\right| )=-\ln \left( \frac{\left|
\mathbf{r}-\mathbf{r}^{\prime }\right| k_{\min }}{2e^{-C }}\right)
=-\ln \frac{\left| \mathbf{r}- \mathbf{r}^{\prime }\right| }{L}$; then
\begin{equation}
\label{eq:kmin}
k_{\min }=\frac{2e^{-C }}{L}\,.
\end{equation}
We note that for the above calculation being consistent it is
necessary to choose $L\to \infty$, as it has been discussed earlier in
the preceding section. The necessity of this choice for the arbitrary
constant $L$ is also discussed the appendix B of Ref.~\cite{jancov l
inf}.

Returning to the calculation of the grand potential in two dimensions,
we impose the infrared cutoff $k_{\min}=\frac{2e^{-C }}{L}$ to
the integral~(\ref{eq:grand-potential-bulk}) to obtain the result for
the grand potential
\begin{equation}
  \label{eq:Omega-bulk-2D}
  \frac{\beta\Omega}{V}=\frac{\kappa^2}{4\pi}\left[-\ln\frac{\kappa
  L}{2}-C+\frac{1}{2}\right]-\sum_{\alpha} \zeta_{\alpha}
\,.
\end{equation}
In the above expression all terms that vanish when $L\to\infty$ have
been omitted. The density--fugacity relation is now
\begin{equation}
  \label{eq:n--z-2D}
  n_{\alpha}=-\zeta_{\alpha}
  \frac{\partial (\beta\Omega/V)}{\partial \zeta_{\alpha}}
  = \zeta_{\alpha}- \zeta_{\alpha}\frac{\beta q_{\alpha}^{2}}{2}
  \left[\ln\frac{\kappa L}{2}+C\right]
\,.
\end{equation}
For a two-component plasma it can be checked that this result is
reproduced from the small-$\beta q^2$ expansion of the exact relation
between the density and the fugacity~\cite{Samaj-TCP}. Notice again
that $\kappa$ can be replaced by $\kappa_{\mathrm{DH}}$ at the order
of approximation we are working on, since
$\kappa=\kappa_{\mathrm{DH}}[1+\mathcal{O}(\Gamma_{2,\alpha} \ln
\Gamma_{2,\alpha})]$. Reporting equation~(\ref{eq:n--z-2D}) back into
equation~(\ref{eq:Omega-bulk-2D}) one obtains the equation of state,
which turns out to be exact at the level of the Debye--H\"uckel
approximation,
\begin{equation}
  \beta p=\sum_{\alpha} n_{\alpha}\left(1-\frac{\beta
  q_{\alpha}^2}{4}\right)
\,.
\end{equation}
Doing the usual Legendre transform $F=\Omega+\sum_{\alpha} \mu_\alpha
N_{\alpha}$, one can recover the known expression for the excess free
energy in the Debye--H\"uckel approximation~\cite{Deutsch,Deutsch-Lavaud}
\begin{equation}\label{eq:Free-Energy-2D-bulk}
  \frac{\beta F_{\mathrm{exc}}}{V}=\frac{\kappa_{\mathrm{DH}}^2}{4\pi}
  \left[\frac{1}{2}-\ln\frac{\kappa_{\mathrm{DH}} L}{2}-C\right] \,.
\end{equation}

To conclude with the results for a two-dimensional system let us
clarify a point regarding the limit $L\to\infty$. Actually in
equation~(\ref{eq:kmin-introduccion}) and below we require that $L$ be
large compared to the average distance between particles which is of
order $n^{-1/2}$ with $n$ the density. In the Debye--H\"uckel
approximation the density $n$ is of the same order as the fugacity
$\zeta$. Therefore we require that $L\zeta^{1/2} \gg 1$. In the
results for the grand-potential~(\ref{eq:Omega-bulk-2D}), the
densities~(\ref{eq:n--z-2D}) and the free
energy~(\ref{eq:Free-Energy-2D-bulk}) appears the quantity $\kappa L$
which is proportional to $(\beta q^2)^{1/2} L\zeta^{1/2}=\Gamma^{1/2}
(L\zeta^{1/2})$. Notice that in the above expression $L\zeta^{1/2} \gg
1$ but the coulombic coupling $\Gamma \ll 1$. Therefore we require
that in two dimensions the Debye--H\"uckel limit should be taken with
$\Gamma\to 0$, $L\zeta^{1/2}\to \infty$ but $\Gamma^{1/2} (L\zeta^{1/2})$
should remain of order 1.

\subsection{The disk}
\label{sec:disk}

We now consider a two-dimensional Coulomb fluid confined in a disk of
radius $R$. To apply the method outlined in section~\ref{sec:method},
we first need to compute the eigenvalues of the Laplace operator for
this geometry. Let $\Psi(r,\varphi )=R(r)\Phi (\varphi ),$ we look for
the solution of the equation $\Delta\Psi(r,\varphi )=\lambda
\Psi(r,\varphi )$. Using the explicit form of the Laplace operator in
polar coordinates we find $\Psi(r,\varphi )=R(r)\Phi (\varphi
)\varpropto e^{\pm il\varphi }I_{l}(\sqrt{\lambda}r)$ where $I_{l}(x)$
is the modified Bessel function of order $l$. Using the boundary
conditions $ \Psi(R,\varphi )=0$; $\Psi(r,0)=\Psi(r,2\pi ),$ we find
$l\in \mathbb{Z}$ and $ I_{l}\left( \sqrt{\lambda _{k}}R\right) =0,$
that is $\sqrt{\lambda _{k}} R=\nu _{l,n}$ is the $n$-th zero of
$I_{l}$.%
\footnote{ Notice that since the zeros of $I_l$ are imaginary then
  $\sqrt{\lambda_k}$ is imaginary, this is expected since the
  Laplacian eigenvalues $\lambda_k$ are negative.}
Then replacing these eigenvalues into equation~(\ref{gp})
gives for the grand potential the expression
\begin{equation}
\beta \Omega =\frac{1}{2}\sum_{l=-\infty }^{\infty }\ln \prod_{n=1}^{\infty}
\left( 1-\frac{R^{2}\kappa^{2}}{\nu _{l,n}^{2}}\right) -
\frac{V\kappa^{2}}{2\left(2\pi \right)}
\int_{k_{\min }}^{\infty}\frac{dk
}{k}-\sum_{\alpha }V\zeta _{\alpha }
\,.
\end{equation}
The second term, written as an integral over $k$, comes from the terms
involving the Laplacian eigenvalues for a non-confined system:
$e^{\kappa^2/\lambda_k^{0}}$, with the same infrared cutoff $k_{\min}$
discussed previously and given by equation~(\ref{eq:kmin}). Both the
sum and the integral diverge separately for large values of $l$ and
$k$ but when we put together both terms the ultraviolet divergences
should cancel. It is however more convenient to compute separately
each term. Therefore we will impose an upper cutoff $N$ for $l$ and
$k_{\max}$ for $k$. Both cutoffs are of course proportional, the exact
relation between $N$ and $k_{\max}$ can be obtained at the end of the
calculations by imposing that for the bulk grand potential we should
recover the result~(\ref{eq:Omega-bulk-2D}) from last section.

Using the infinite product representation of the modified Bessel
function $ \prod_{n=1}\left( 1-\frac{\kappa^{2} R^2}{\nu
_{l,n}^{2}}\right) =l!\left( \frac{2}{\kappa R }\right)
^{l}I_{l}(\kappa R)$, \cite{abramovitz} and the property
$I_{l}(x)=I_{-l}(x)$ we have
\begin{eqnarray}
\beta \Omega &=&\sum_{l=0}^{N}\ln l!+
\ln \left( \frac{2}{\kappa R}\right)
\sum_{l=0}^{N }l
+\sum_{l=0}^{N }\ln I_{l}(\kappa R)-\frac{1}{2}\ln \left[
I_{0}(\kappa R)\right]\\
&&
-\frac{\kappa^{2}V}{4\pi }\int_{k_{\min }}^{k_{\max }}\frac{dk}{k
}-\sum_{\alpha }V\zeta _{\alpha }  
\,.
\label{omega}
\end{eqnarray}
Using Stirling approximation: $\ln N!=N\ln N-N+\frac{1}{2}\ln (2\pi
N)+\cdots $, Euler-McLaurin summation formula:
$\sum_{l=0}^{N}f(l)=\int_{0}^{N}f(l)dl+\frac{1}{2}\left[ f(0)+f(N)
\right] +\frac{1}{12}\left[ f^{\prime }(N)-f^{\prime }(0)\right]
+\cdots $ and the uniform Debye expansion~\cite{abramovitz} for $\ln
I_{l}(z )$, valid for large $z$,
\begin{equation}
\label{eq:Debye-ex-I_l}
\fl
\ln I_{l}(z )=-\frac{1}{2}\ln (2\pi )-\frac{1}{4}\ln \left( z
^{2}+l^{2}\right) +\eta (l,z )+\frac{3u-5u^{3}}{24l}+o\left( \frac{1}{
z ^{2}+l^{2}}\right)
\end{equation}
\begin{equation}
\label{eq:params-Debye-exp}
\eta (l,z)=\left( z ^{2}+l^{2}\right) ^{1/2}-l\sinh ^{-1}\left( 
\frac{l}{z }\right) ;\quad u=\frac{l}{\left( z
^{2}+l^{2}\right) ^{1/2}}
\end{equation}
after some calculations we finally obtain the large-$R$ expansion
\begin{equation}
\fl
\beta \Omega =\frac{\kappa^{2}R^{2}}{4}
\left( 1+\ln \frac{2e^{-C }}{\kappa L}\frac{
2N}{k_{\max }R}-
\frac{4\pi\sum_{\alpha} \zeta_{\alpha} }{\kappa^{2}}\right) 
-R\left( \frac{\kappa\pi }{4}
\right) +\frac{1}{6}\ln R+\mathcal{O}(R^{0})  
\,.
\label{fin}
\end{equation}
In the above expression all terms that vanish when $N\to\infty$ and
$k_{\max}\to\infty$ have been omitted. The bulk term (proportional to
$\pi R^2$) of the above equation~(\ref {fin}) should be the same as in
equation~(\ref{eq:Omega-bulk-2D}) therefore the cutoffs $N$ and
$k_{\max}$ should be related by $k_{\max
}=\frac{2N}{R}e^{1/2}$. Finally
\begin{equation}
\beta \Omega = \beta \omega_b \pi R^2 +2\pi R\beta \gamma
+  \frac{1}{6}\ln (\kappa R)+\mathcal{O}(R^{0})  
\label{expansiondisco}
\end{equation}
with the bulk grand potential per unit volume $\omega_b$ (equal to
minus the bulk pressure $p_b$) given by
\begin{equation}
  \label{eq:omega-bulk}
  \beta\omega_b=-\beta
  p_{b}=\frac{\kappa^2}{4\pi}\left[-\ln\frac{\kappa
  L}{2}-C+\frac{1}{2}\right]-\sum_{\alpha} \zeta_{\alpha}
\end{equation}
and the surface tension $\gamma$ is given by
\begin{equation}
  \label{eq:surface-tension}
  \beta \gamma = -\frac{\kappa}{8}
\,.
\end{equation}
The two-dimensional two-component plasma near a plane ideal conductor
electrode has been solved exactly~\cite{Samaj-Janco-TCP-metal}. For a
two-component plasma our result~(\ref{eq:surface-tension}) for the
surface tension agrees with the lower order expansion in $\beta q^2$
of the exact result of Ref.~\cite{Samaj-Janco-TCP-metal}. In
equation~(\ref{expansiondisco}) we notice the existence of the
universal logarithmic finite-size correction $(1/6)\ln R$ with
$\chi=1$ for the disk.

\subsection{The annulus}

We now consider a Coulomb fluid confined in an annulus of inner radius
$R_1$ and outer radius $R_2$. As before we need to calculate the
eigenvalues of the Laplace operator for this geometry. The
eigenfunction of the Laplacian with eigenvalue $\lambda$, in this
geometry, is $\Psi(r,\varphi )=\left[ AI_{l}(\sqrt{ \lambda
}r)+BK_{l}(\sqrt{\lambda }r)\right] e^{il\varphi }$. Imposing the
Dirichlet boundary conditions yields the linear system of equations
$\Psi(R_1,\varphi )=0$ and $\Psi(R_2,\varphi )=0$. To have a
non-vanishing solution for the eigenproblem we require that the
determinant of this system vanishes. This gives the equation that
defines the eigenvalues for this problem,
\begin{equation}
I_{l}(\sqrt{\lambda }R_1)K_{l}(\sqrt{\lambda }R_2)-K_{l}(\sqrt{\lambda
}R_1)I_{l}(\sqrt{\lambda }R_2)=0 
\label{equ JN-JN}
\end{equation}
this means that $\lambda _{k}=z_{l,n}^{2}$ where $z_{l,n}$
is the $n$-th root of equation 
\begin{equation}
\label{eq:eigenvals}
I_{l}(zR_1)K_{l}(zR_2)-K_{l}(zR_1)I_{l}(zR_2)=0   \,.
\end{equation}
Notice that the roots of this equation are the same for $l$ and $-l$,
therefore we will concentrate on the case $l>0$.  To compute the grand
partition function from equation~(\ref{gp}) we need to evaluate the
infinite product
$\prod_{l}\prod_{n}\left(1-\frac{\kappa^2}{z_{l,n}^{2}}\right)$. For a
given $l$, the product over the index $n$ of the roots of
equation~(\ref{eq:eigenvals}) can be performed using a generalization
of the infinite product representation of the Bessel functions used in
the last section~\cite{Finite zise Gabriel, Tellez-tcp-disque-neumann,
Forrester-JSP}. For $l>0$, let us introduce the entire function
\begin{equation}
  f_{l}(z)= \frac{2l}{\left(\frac{R_1}{R_2}\right)^{l}
    -\left(\frac{R_2}{R_1}\right)^{l}}
  \left[
    I_{l}(zR_1)K_{l}(zR_2)-K_{l}(zR_1)I_{l}(zR_2)
    \right]
\,.
\end{equation}
By construction the zeros of the function $f_{l}$ are $z_{l,n}$ and it
has the following properties: $f_l(0)=1$, $f'_l(0)=0$ and
$f_l(z)=f_l(-z)$. Therefore $f_l$ admits a Weierstrass infinite product
representation of the form~\cite{Whittaker-Watson}
\begin{equation}
  f_l(z)=\prod_{n} \left(1-\frac{z^2}{z^2_{l,n}}\right)
\,.
\end{equation}
Then the infinite product we wish to evaluate is simply
$\prod_{n}(1-\kappa^2/z_{l,n}^{2})=f_l(\kappa)$. For $l=0$ the
function $f_0$ should read
\begin{equation}
  f_0(z)=\frac{1}{\ln(R_1/R_2)}
  \left[
    I_{0}(zR_1)K_{0}(zR_2)-K_{0}(zR_1)I_{0}(zR_2)
    \right]
  \,.
\end{equation}
The grand potential is then given by
\begin{equation}
  \label{eq:gp-annulus}
  \beta\Omega=\sum_{l=1}^{N} \ln f_{l}(\kappa)
  +\frac{1}{2} \ln f_{0}(\kappa) 
  -\frac{\kappa ^{2}V}{4\pi }\int_{k_{\min }}^{k_{\max }}\frac{dk}{k}%
  -\sum_{\alpha }V\zeta _{\alpha }  \,.
\end{equation}
As in the case of the disk we regularize the summation on $l$ by
introducing an upper cutoff $N$ and the integral with an ultraviolet
cutoff $k_{\max}$. These cutoffs are proportional in order to cancel
the divergences. However their exact relationship is a priori
different from the one in the disk case and can be found at the end of
the calculations by requiring that we recover the same bulk value of
the grand potential as in the previous examples. On the other hand the
infrared cutoff $k_{\min }=\frac{2e^{-C }}{L}$ is the same as before.

We now proceed to find the finite-size expansion of the grand
potential. We consider a very large annulus with $R_1\to\infty$,
$R_2\to\infty$, $R_2-R_1\to\infty$ and $x=R_1/R_2<1$ finite and
fixed. The calculations are similar to those of the disk, we now use
the uniform Debye expansion of $\ln K_{l}(z)$ valid for large
arguments~\cite{abramovitz}
\begin{equation}
\fl
\ln K_{l}(z)=\ln \left[ \frac{\sqrt{\pi }}{\sqrt{2}}\right]
-\frac{1}{4}\ln (l^{2}+z^{2})-\eta (l,z)+\ln \left[
1-\frac{3u-5u^{3}}{24l}\right] +o\left(\frac{1
}{l^{2}+z^{2}}
\right)
\end{equation}
with $\eta (l,z)$ and $u$ defined in
equation~(\ref{eq:params-Debye-exp}). Notice that in the functions
$f_l(\kappa)$, the contribution of $K_{l}(\kappa R_2)I_{l}(\kappa
R_1)$ is exponentially smaller than the one from the term
$I_{l}(\kappa R_2)K_{l}(\kappa R_1)$. Using again the Euler-McLaurin
summation formula to transform the sum over $l$ into an integral,
after some calculations we find in the limit $N\to\infty$,
\begin{eqnarray}
  \label{eq:gp-annulus-1}
  \beta \Omega&=&
  \frac{1}{4}\left(R_2^2\ln\frac{2N}{k_{\max}R_2}
  -R_1^2\ln\frac{2N}{k_{\max}R_1}\right)\\
  &+&
  \frac{\kappa^2(R_2^2-R_1^2)}{4}
  \left(\frac{1}{2}+\ln\frac{2e^{-C}}{\kappa L}\right)
  -\frac{\pi}{4}\kappa(R_2+R_1)
  +\mathcal{O}(1)
\,.
\end{eqnarray}
All terms that vanish when $N\to\infty$ have been omitted. To recover
the proper bulk value of the grand potential and ensure extensivity,
the first term in equation~(\ref{eq:gp-annulus-1}) should vanish. This
imposes the relationship between the ultraviolet cutoffs $N$ and
$k_{\max}$,
\begin{equation}
  \frac{2Ne^{1/2}}{k_{\max}}=R_2 x^{\frac{x^2}{x^2-1}}
\,.
\end{equation}
This relation is similar to the one found in the disk replacing $R$ by
$R_2 x^{\frac{x^2}{x^2-1}}$.

Returning to the grand potential we conclude that its the large annulus
expansion is
\begin{equation}
  \Omega=\pi (R_2^2-R_1^2) \omega_b+ 2\pi(R_1+R_2)\gamma + \mathcal{O}(1)
\end{equation}
with the $\omega_b$ and $\gamma$ given by
equations~(\ref{eq:omega-bulk}) and~(\ref{eq:surface-tension})
respectively.  In the $\mathcal{O}(1)$ neglected terms there are terms
of the form $\ln(R_1/R_2)$ and more generally functions of $x=R_1/R_2$
which are indeed of order 1. There are not logarithmic finite-size
corrections, such as $\ln (\kappa \sqrt{R_1 R_2})$, according to the
fact that $\chi =0$ for an annulus.

\section{Summary and conclusion}

The method presented here gives a practical prescription for the
calculation of finite-size corrections of the grand potential of a
Coulomb system in the Debye--H\"uckel regime, that can be easily
applied to more complicated geometries in two and three
dimensions. The proper substraction of the self-energies avoids the
divergence of the infinite products involved in the calculations.  In
the disk and annulus geometry that we used to illustrate our method,
we recovered the bulk pressure and the surface tension of the system
in the Debye--H\"uckel regime. For the disk we obtained a universal
finite-size correction $\frac{1}{6}\ln R$, with the expected value
$\chi =1$, for the Euler characteristic of the disk. For the annulus
since $\chi=0$ no finite-size correction is expected and we confirmed
this result by direct calculation of the finite-size expansion. In the
case of a system in a domain of arbitrary shape, the logarithmic
universal correction to the grand potential may be obtained from the
asymptotic properties of the spectrum of the Laplace operator and its
relation with the geometry of the manifold for which this spectrum is
calculated~\cite{Kac, curvature eigenv laplacian}. Work in this
direction is in progress.

\ack We wish to thank the Fondo de Investigaciones de la Facultad de
Ciencias de la Universidad de Los Andes, the Banco de la Rep\'ublica
de Colombia and ECOS-Nord/COLCIENCIAS--ICFES--ICETEX for their
financial support in the development of this work.  G.~T. thanks
B.~Jancovici for some useful comments and discussions, and for a
careful reading of the manuscript. We thank also the referees for
useful remarks, some of which lead to development
of~\ref{sec:appendix-B}.

\appendix

\section{Relationship with the usual formulation of Debye--H\"uckel theory}
\label{sec:appendix}

The usual formulation of Debye--H\"uckel theory~\cite{McQuarrie}, for
a confined Coulomb system with Dirichlet boundary conditions for the
electric potential, starts by computing the electric potential
$\Phi_{\alpha}(\mathbf{r},\mathbf{r}')$ created at $\mathbf{r}'$ by a
particle of charge $q_{\alpha}$ located at $\mathbf{r}$ and its
polarization cloud. We have
$\Phi_{\alpha}(\mathbf{r},\mathbf{r}')=q_{\alpha} K(\mathbf{r},
\mathbf{r}')$ with the Debye--H\"uckel kernel $K$ that satisfies
\begin{equation}
  \left(\Delta - \kappa_{\mathrm{DH}}^2\right)
  K(\mathbf{r},\mathbf{r}')= -s_d \delta(\mathbf{r}-\mathbf{r}')
\end{equation}
with $\kappa_{\mathrm{DH}}^2=\sum_{\alpha} \beta q_{\alpha}^{2}
n_{\alpha} s_d$. Formally $K$ can be written as
\begin{equation}
  K(\mathbf{r},\mathbf{r}')=
  \left\langle \mathbf{r} \left|
  \frac{-s_d}{\Delta - \kappa_{\mathrm{DH}}^2}
  \right|\mathbf{r}'\right\rangle
\end{equation}
where the Laplacian is considered to satisfy Dirichlet boundary
conditions. Then, one computes the internal
potential energy $U$ of the system as
\begin{equation}
  U=\frac{1}{2} \int d\mathbf{r} \sum_{\alpha}
  q_{\alpha}^2 n_{\alpha} 
  \lim_{\mathbf{r}'\to\mathbf{r}}
  \left(
  K(\mathbf{r},\mathbf{r}')-v_d^{0}(\mathbf{r},\mathbf{r}')
  \right)
\end{equation}
which can formally be written as
\begin{eqnarray}
  U&=&- \frac{1}{2}  \sum_{\alpha} q_{\alpha}^2
  n_{\alpha} s_d
  \int d\mathbf{r} \left\langle\mathbf{r} \left|
  \frac{1}{\Delta - \kappa_{\mathrm{DH}}^2} - 
  \frac{1}{\Delta^{0}}
  \right|\mathbf{r}\right\rangle
  \\
  &=& \frac{\kappa_{\mathrm{DH}}^2}{2\beta} 
  \Tr\left[
    \frac{1}{\Delta - \kappa_{\mathrm{DH}}^2} - 
    \frac{1}{\Delta^{0}}
    \right]
  \\
  &=&
  \frac{\kappa_{\mathrm{DH}}^2}{2\beta}  
  \sum_{n} \left[
  \frac{1}{\lambda_n - \kappa_{\mathrm{DH}}^2} - 
  \frac{1}{\lambda_{n}^{0}}
  \right]
  \,.
  \label{eq:U-direct}
\end{eqnarray}
The notation $\Delta^0$ denotes the Laplacian operator with free
boundary conditions. 

On the other hand the internal excess energy $U$ can be computed from
the thermodynamic relation $U=-\left(\partial\ln\Xi/\partial \beta
\right)_{\zeta,V}$. Using the sine-Gordon formulation, we can obtain
an independent expression for the internal excess energy and compare
it to equation~(\ref{eq:U-direct}). Using equation~(\ref{gp}) gives
\begin{eqnarray}
  \label{eq:U-gp}
  -\left(
  \frac{\partial\ln\Xi}{\partial\beta}
  \right)_{\zeta,V}
  &=&
  \frac{1}{2}
  \frac{\partial}{\partial\beta}
  \sum_{n} \left[\ln\left(1-\frac{\kappa^2}{\lambda_n}\right)
    + \frac{\kappa^2}{\lambda_n^{0}}\right]
  \nonumber\\
  &=&
  \frac{\kappa^2}{2\beta} 
  \sum_{n} \left[
  \frac{1}{\lambda_n - \kappa^2} - 
  \frac{1}{\lambda_{n}^{0}}
  \right]
  \,.
\end{eqnarray}
At the Debye--H\"uckel level of approximation $\kappa_{\mathrm{DH}}$
(expressed in terms of the densities) can be replaced by $\kappa$
(expressed in term of the fugacities) with corrections of higher
order. Therefore with equation~(\ref{gp}) for the grand potential and
equation~(\ref{eq:U-gp}), one recovers the
expression~(\ref{eq:U-direct}) for the internal excess energy obtained
from the usual formulation of Debye--H\"uckel theory.

\section{On the pseudo-neutrality condition and the potential
  difference between the system and the walls}
\label{sec:appendix-B}

Coulomb systems have the interesting property that any excess charge
in the system is expelled to the
boundaries~\cite{Lieb-Lebowitz}. Therefore any infinite system is
neutral. When the system is described in the grand canonical ensemble
with fugacities $\zeta_{\alpha}^{*}$ the electroneutrality has the
consequence that the fugacities are not independent. Several choices
of the fugacities can describe the same system. More precisely, the
grand potential does not depend on the combination $\sum_{\alpha}
q_{\alpha}
\zeta_{\alpha}^{*}$~\cite{review,Lieb-Lebowitz,Aqua-Cornu}. Therefore
one can impose the so-called pseudo-neutrality condition
\begin{equation}
\label{eq:pseudo-neutrality}
\sum_{\alpha} q_{\alpha} \zeta_{\alpha}^{*}=0
\,.
\end{equation}

For a confined Coulomb system the situation is more involved. Suppose
that the confined system, described in the grand canonical ensemble
with fugacities $\zeta_{\alpha}$, is in equilibrium with an infinite
neutral reservoir at zero electric potential with fugacities
$\zeta_{\alpha}^{*}$ that satisfy the pseudo-neutrality
condition~(\ref{eq:pseudo-neutrality}). Let us consider that the
confined system is large and that far from the boundaries the average
electric potential of the system is a constant $\psi_{0}$. Writing
down the equilibrium condition that the electrochemical potentials of
the system and the reservoir should be equal yields
$\zeta_{\alpha}^{*}=\zeta_{\alpha} e^{-\beta q_{\alpha}
\psi_0}$. Therefore the confined system can be described with the
fugacities $\zeta_{\alpha}$ which a priori do not satisfy the
pseudo-neutrality condition or with the fugacities
$\zeta_{\alpha}^{*}$ which satisfy the pseudo-neutrality condition
plus the parameter $\psi_0$ which is the potential difference between
the system and the reservoir. In this article we have supposed so far
that the fugacities $\zeta_{\alpha}$ satisfy the pseudo-neutrality
conditions. In the following we will consider the general case when
the fugacities do not satisfy the pseudo-neutrality condition and we
will explore how they are related to the potential $\psi_0\neq 0$ in
an approximate mean field picture. If $\psi_0\neq 0$ this potential
difference will create a surface charge density near the boundaries
and we would expect that this effect will add to the grand potential
and the free energy a \textit{surface} term. We will show that this
contribution turn out to be $-(1/2) Q \psi_0$ where $Q$ is the excess
charge of the system which is spread over the surface of the
boundaries~\cite{Lieb-Lebowitz}.

Actually we can justify this argument within our formalism of the
sine-Gordon transformation by adapting some arguments put forward in
Ref.~\cite{review} for the case of free boundary conditions. In the
case of two dimensional systems, we will also show that if the
potential difference $\psi_0$ is not too high, the contributed surface
term is of higher order in the coulombic coupling constant than the
surface tension already computed in section~\ref{sec:disk} and given
by equation~(\ref{eq:surface-tension}) and therefore it can be
neglected in the Debye--H\"uckel approximation.

Let us rewrite equation~(\ref{G}) for the grand canonical partition
function as
\begin{equation}
  \Xi(\beta,\zeta_{\alpha},V)=
  \frac{1}{Z} 
  \int \mathcal{D}\phi\,
  \exp[-S]
\end{equation}
with the action $S$ given by 
\begin{equation}
  \label{eq:action}
  -S=\int \left[
    \frac{\beta}{2s_d} \phi \Delta \phi + \sum_{\alpha} \zeta_{\alpha}
    e^{-i \beta q_{\alpha} \phi}\right]
  \,.
\end{equation}
To lighten the notation, in this appendix we will often omit the
variable $\mathbf{r}$ in the integrals: $\int \phi = \int
\phi(\mathbf{r})\,d\mathbf{r}$. For simplicity we have omitted the
self-energy term which is irrelevant in the present discussion (one
could eventually consider that the fugacities $\zeta_{\alpha}$ are
renormalized by a multiplication by $e^{\beta q_{\alpha}^{2}
v_{S-E}/2}$). 

If the fugacities do not satisfy the pseudo-neutrality condition, the
stationary point of the action $S$ is not $\phi=0$ as before. Let
$\psi(\mathbf{r})$ be $i$ times the solution of $\delta S/\delta
\phi=0$. This field satisfies
\begin{equation}\label{eq:PB}
  \Delta \psi(\mathbf{r}) +
  s_d \sum_{\alpha} \zeta_{\alpha} q_{\alpha} e^{-\beta q_{\alpha}
  \psi(\mathbf{r})} = 0
  \,.
\end{equation}
with Dirichlet boundary conditions: $\psi(\mathbf{r})$ vanishes on the
boundary. This is Poisson--Boltzmann equation and the field
$\psi(\mathbf{r})$ is the average electrostatic potential in the mean
field approximation~\cite{Kennedy-mean-field}. Notice that if the
fugacities $\zeta_{\alpha}$ satisfy the pseudo-neutrality condition
then $\psi(\mathbf{r})=0$ is a solution of Poisson--Boltzmann
equation~(\ref{eq:PB}). In this case, and in the mean field
approximation, the potential difference between the boundaries and the
system is zero. If the fugacities $\zeta_{\alpha}$ do not satisfy the
pseudo-neutrality condition then $\psi(\mathbf{r})$ is not zero and
contrary to what has been done before the expansion to the quadratic
order of the action $S$ should now be done around $\phi=-i\psi$
instead of $\phi=0$. To accomplish this let us do the change of
variable in the functional integral $\phi'=\phi+i\psi$. We have
$\mathcal{D}\phi=\mathcal{D}\phi'$ and the action is now given by
\begin{equation}
  -S=\int \frac{\beta}{2s_d} \left[
    \phi'\Delta \phi' -2i \phi'\Delta \psi - \psi\Delta\psi\right]
    + \sum_{\alpha}
    \zeta_{\alpha} e^{-\beta q_{\alpha} (\psi+i\phi')}
    \,.   
\end{equation}
The new field $\phi'$ fluctuates around $0$ and now we can expand the
exponential to the second order in the coulombic coupling. The linear
terms in $\phi'$ in the action $S$ are canceled by applying the
stationary condition (Poisson--Boltzmann) equation~(\ref{eq:PB}) and
we find $S=S_{1}+S_{2}+o(\Gamma_{d,\alpha}^{d/2})$ with
\begin{eqnarray}
  \label{eq:S1-init}
  S_{1}&=&
  \frac{1}{2}\int \phi'(\mathbf{r})\left(
  \frac{-\beta \Delta}{s_d} +
  (\beta q_{\alpha})^2 \zeta_{\alpha} e^{-\beta q_{\alpha}\psi(\mathbf{r})}
  \right)\phi'(\mathbf{r})\, d\mathbf{r}
  \\
  \label{eq:S2-init}
  S_{2}
  &=&
  \int\left[
  \frac{\beta}{2s_d} \psi(\mathbf{r}) \Delta \psi(\mathbf{r})
  -\sum_{\alpha}\zeta_{\alpha} e^{-\beta q_{\alpha} \psi(\mathbf{r})}
  \right] \, d\mathbf{r}
\,.
\end{eqnarray}
The term $S_1$ is of order $\Gamma_{d,\alpha}^{d/2}$ in the coupling
constants. To verify this, note that in two dimensions the field
$\phi'$ can be written as $q f(\kappa \mathbf{r})$ with $f$ some
function of order one and $q$ is the magnitude of the elementary
charges in the system, for example $q=\max |q_{\alpha}|$. Rescaling
the distances in the integral by the inverse Debye length $\kappa$
shows that $S_1$ is of order $\Gamma_{2,\alpha}$. In three dimensions
$\phi'=q\kappa f(\kappa \mathbf{r})$ and doing the same scaling in the
integral as above shows that $S_1$ is now of order
$\Gamma_{3,\alpha}^{3/2}$. To know the order of magnitude of $S_2$ we
need further assumptions. To proceed, we shall need in principle the
solution $\psi(\mathbf{r})$ of Poisson--Boltzmann
equation~(\ref{eq:PB}). However the solution of Poisson--Boltzmann
equation is not known explicitly except for a few very simple
geometries~\cite{Gouy,Chapman,Grahame,TracyWidom}. Nevertheless the
qualitative behavior of the mean field $\psi(\mathbf{r})$ is very
simple. It vanishes on the boundary and a few screening lengths away
from the boundary it is almost equal to a constant value
$\psi_0$. This constant average value of the potential $\psi_0$ is
given by Poisson--Boltzmann equation~(\ref{eq:PB}) for a constant
field:
\begin{equation}\label{eq:pseudo-neutral-2}
  \sum_{\alpha} q_{\alpha} \zeta_{\alpha} e^{-\beta
  q_{\alpha}\psi_0}=0
  \,.
\end{equation}
Let us define the renormalized fugacities
$\zeta_{\alpha}^{*}=\zeta_{\alpha} e^{-\beta q_{\alpha}\psi_0}$. By
equation~(\ref{eq:pseudo-neutral-2}) these new fugacities satisfy the
pseudo-neutrality condition~(\ref{eq:pseudo-neutrality}). The physical
interpretation of these new fugacities is of course the one exposed at
the beginning of this appendix: they are the fugacities of the
infinite neutral grounded reservoir. Let us now write
$\psi(\mathbf{r})=\psi_0+\delta\psi(\mathbf{r})$. Notice that
$\delta\psi(\mathbf{r})$ is almost zero in the deep interior of the
system and only has significant values near the boundaries. Let us
suppose that the variations of $\delta\psi(\mathbf{r})$ are small,
more precisely let us suppose that $\delta\psi(\mathbf{r})=q g(\kappa
\mathbf{r})$ in two dimensions or $\delta\psi(\mathbf{r})=\kappa q
g(\kappa \mathbf{r})$ in three dimensions with $g$ some function of
order one. Then expanding $S_1$ for small $\delta\psi$ yields
\begin{equation}
  \label{eq:S1-expand}
  S_1= \frac{1}{2}\int \phi'(\mathbf{r})\left( \frac{-\beta
  \Delta}{s_d} + (\beta q_{\alpha})^2 \zeta_{\alpha}^{*}
  \right)\phi'(\mathbf{r})\, d\mathbf{r}
  +\mathcal{O}(\Gamma_{d,\alpha}^{d})
  \,.
\end{equation}
For the second part of the action the same expansion yields
\begin{equation}
  \fl
  \label{eq:S2-partial}
  S_2=-\sum_{\alpha} \zeta_{\alpha}^{*} V
  +\frac{\beta \psi_0}{2s_d}\int \Delta(\delta\psi)
  +\frac{\beta}{2s_d} \int \delta\psi (\Delta - \kappa^2) \delta\psi
  +\mathcal{O}(\Gamma_{d,\alpha}^{d})
\end{equation}
where we have know defined the inverse Debye length $\kappa$ in terms
of the renormalized fugacities $\zeta_{\alpha}^{*}$ as
$\kappa=\sqrt{\beta s_d\sum_{\alpha}\zeta_{\alpha}^{*} q_{\alpha}^2}$.
Actually a closer inspection of equation~(\ref{eq:S2-partial}) shows
that the last term of $S_2$ is actually of higher order that the two
other terms. Indeed expanding Poisson--Boltzmann
equation~(\ref{eq:PB}) for $\delta\psi$ small shows that
\begin{equation}
  \label{eq:DH-lineal}
  \Delta(\delta\psi)-\kappa^2\delta\psi=
  \mathcal{O}\left(\zeta_{\alpha}^{*} q_{\alpha} 
  (\beta q_{\alpha}\delta\psi)^2\right)
\,.
\end{equation}
Therefore
\begin{equation}
  -\frac{\beta}{2s_d} \int \delta\psi (\Delta - \kappa^2) \delta\psi
  =\mathcal{O}(\Gamma_{d,\alpha}^d)
\end{equation}
and 
\begin{equation}
  \label{eq:S2}
  S_2=-\sum_{\alpha} \zeta_{\alpha}^{*} V
  +\frac{\beta \psi_0}{2s_d}\int_{V} \Delta(\delta\psi)  
  +\mathcal{O}(\Gamma_{d,\alpha}^{d})
  \,.
\end{equation}
Notice now that only $S_1$ depends on $\phi'$ and the result of the
functional Gaussian integration over the field $\phi'$ will be the
same as in section~\ref{sec:method},
equation~(\ref{eq:gaussian-inte}), except that the fugacities
$\zeta_{\alpha}$ have to be replaced by $\zeta_{\alpha}^{*}$. The term
$S_2$ does not depend on the field $\phi'$ and will give only an
additive contribution to the grand potential. Finally we obtain for
the grand potential
\begin{equation}
  \Omega=\frac{k_B T}{2}\ln \left(
  \prod_{m}\left( 1-\frac{\kappa^{2}}{\lambda _{m}}\right)
  \prod_{n} e^{\frac{\kappa^{2}}{\lambda _{n}^{0}}} \right)
  -k_B T\sum_{\alpha }V\zeta^{*}_{\alpha } + \Omega_{S}
    \, .
\end{equation}
That is the same grand potential as before but evaluated for the
fugacities $\zeta_{\alpha}^{*}$ instead of $\zeta_{\alpha}$ plus a
contribution
\begin{equation}\label{eq:surface-contrib-extra}
  \Omega_{S}=S_2+\sum_{\alpha} \zeta_{\alpha}^{*} V=-\frac{
  \psi_0}{2}\int_{V} \rho(\mathbf{r})\,d\mathbf{r} = - \frac{1}{2}
  \psi_0 Q \,.
\end{equation}
We have used Poisson equation $\Delta\psi=-s_d \rho$ with $\rho$ the
average charge density of the system in the mean field
approximation. The excess total charge is $Q=\int_{ V} \rho$. Let us
remark a few points about this term. The charge density
$\rho(\mathbf{r})$ is different from zero near the boundaries and a
few Debye lengths away from the boudaries it vanishes. The charge $Q$
is spread near the surface of the system. Therefore the additional
contribution $\Omega_{S}$ to the grand potential is actually a surface
contribution. This is even more clear if from equation~(\ref{eq:S2})
we use Gauss theorem to write $\Omega_{S}$ as
\begin{equation}\label{eq:surface-contrib-extra-bis}
  \Omega_{S}=\frac{ \psi_0}{2s_d}\oint_{\partial V}
  \nabla\psi(\mathbf{r})\cdot d\mathbf{S} = 
  \frac{ \psi_0}{2}\oint_{\partial V} \sigma_w(\mathbf{r})\,
  dS
\end{equation}
where $\sigma_w=s_d\partial_n \psi$ is the surface charge density
induced in the boundary walls of ideal conductor. This charge is
external to the system. Since the ideal conductor is grounded and it
is in total influence with the Coulomb system we have $\oint_{\partial
V} \sigma_w dS= - Q$, thus recovering
equation~(\ref{eq:surface-contrib-extra}). This term could further be
expressed as
\begin{equation}
  \Omega_{S}=\frac{\psi_0 \kappa^{2}}{2s_d}\int_{V}
  \delta\psi(\mathbf{r}) \,d\mathbf{r}
\end{equation}
where we have used equation~(\ref{eq:DH-lineal}). Let us point out
that this additional contribution $\Omega_S$ is not the naive
electrostatic energy $U_{\mathrm{elst}}=(1/2) \int_{V}
\rho(\mathbf{r}) \psi(\mathbf{r}) \, d\mathbf{r}$. This can be checked
for the two-dimensional two-component plasma near a planar wall made
of ideal conductor. Using the results from
Ref.~\cite{Samaj-Janco-TCP-metal}, for small coulombic coupling, the
mean field electric potential is $\psi(x)=\psi_0(1-e^{-\kappa x})$,
with $x$ the distance from the wall. Then in this case
$U_{\mathrm{elst}}=-\Omega_{S}/2$.

In general in two or three dimensions $\Omega_S$ is of order
$\Gamma_{d,\alpha}$. However for two dimensional systems when the
pseudo-neutrality condition is satisfied, we have found a surface
tension given by equation~(\ref{eq:surface-tension}) which is of lower
order $\Gamma_{2,\alpha}^{1/2}$. Therefore in two dimensions the
surface contribution $\Omega_S$ to the grand potential due to the
potential difference $\psi_0$ between the system and the reservoir
found in this appendix can be neglected in front of the surface
tension given in equation~(\ref{eq:surface-tension}). This fact has
also been noticed in Ref.~\cite{Samaj-Janco-TCP-metal} where the exact
expression of the surface tension for a symmetric two-component plasma
has been computed and its expansion for small coulombic coupling
parameter shows that the potential $\psi_0$ do not contribute in the
dominant order.

For two dimensional systems, we can conclude that the results of
sections~\ref{sec:method} and~\ref{sec:confined} also apply when the
fugacities do not satisfy the pseudo-neutrality condition, provided
that one replaces the original fugacities $\zeta_{\alpha}$ with the
renormalized fugacities $\zeta_{\alpha}^*=\zeta_{\alpha} e^{-\beta
q_{\alpha} \psi_0}$ which do satisfy the pseudo-neutrality condition.

In three dimensions the situation is somehow different. By dimensional
analysis one would expect that for a system of characteristic size $R$
and with $\psi_0=0$ the surface term of the grand potential will be
proportional to $\kappa^2 R^2$, therefore of order $\Gamma_{3,\alpha}$.%
\footnote{\label{footnote}
  Applying our method to three dimensional systems, actually
  gives a surface tension $\gamma=\beta^{-1}
  (\kappa^2/16\pi)\ln(\kappa/k_{\max})$ with $k_{\max}$ an ultraviolet
  cutoff. This result will be reported in a future work currently
  under preparation. Therefore it turns out that the dominant term in
  the surface tension is of order
  $\Gamma_{3,\alpha}\ln\Gamma_{3,\alpha}$ and the correction
  $\Omega_S$ is again of higher order.
}\ %
Then for a system with $\psi_0\neq 0$ the
additional surface contribution $\Omega_S$ found in this appendix will
in principle contribute to the total surface tension (see, however, the
footnote).

Let us mention that if the potential difference $\psi_0$ is large one
should go back to equations~(\ref{eq:S1-init}) and~(\ref{eq:S2-init})
and try to study the whole non-linear problem. There is an interesting
regime where the fluctuations $\phi'$ around the mean field are small
enough to expand the action $S$ to the second order as $S=S_1+S_2$ but
that the mean field $\psi_0$ could be large and the further expansion
of $S_1$, equation~(\ref{eq:S1-expand}), and $S_2$,
equation~(\ref{eq:S2}), is not possible. It is expected that some very
interesting phenomena could occur in this non-linear regime, for
instance renormalization and saturation of the surface charge $Q$ and
the potential $\psi_0$, like in the studies of highly charged
colloids~\cite{Alexander,Bocquet-Trizac-Aubouy-JCP-charge-sat,
Tellez-Trizac-charge-sat-PRE}.

To conclude this discussion we should point out a delicate point. The
analysis done in this appendix is based on a mean field approximation:
the function $\psi(\mathbf{r})$ and its constant value $\psi_0$ inside
the system are the solution of Poisson--Bolzmann
equation~(\ref{eq:PB}). In full generality they are different from the
average electric potential inside the system. Only at the first order,
in the ideal gas approximation ($n_{\alpha}=\zeta_{\alpha}^{*}$) we
can identify both. Nevertheless our goal was to obtain an estimation
of the corrections to the grand-potential when the pseudo-neutrality
condition is not satisfied, and based on this estimate we can conclude
that these corrections are of higher order.

\end{document}